\documentclass[aps,pra,showpacs,reprint,superscriptaddress,twocolumn]{revtex4-1}
 \usepackage{graphicx}
 \usepackage{subfigure}
 \usepackage{dcolumn}
 \usepackage{bm}
 \usepackage{sidecap} 
 \usepackage{amsmath}
 \usepackage{amssymb}
 \usepackage{color}

 \usepackage{dsfont}
 \usepackage{units}
 \usepackage{placeins}
 \usepackage{braket}


\begin{document}

\title{Electrically-tunable hole g-factor of an optically-active quantum dot for fast spin rotations} 

\author{Jonathan H. Prechtel}
\email[]{jonathan.prechtel@unibas.ch}
\homepage[]{http://nano-photonics.unibas.ch/}
\author{Franziska Maier}
\affiliation{Department of Physics, University of Basel, Klingelbergstrasse 82, CH-4056 Basel, Switzerland}

\author{Julien Houel}
\affiliation{Institut Lumiere Matiere, CNRS UMR5306, Universite Lyon 1, 69622 Villeurbanne, Cedex, France}

\author{Andreas V. Kuhlmann}
\affiliation{Department of Physics, University of Basel, Klingelbergstrasse 82, CH-4056 Basel, Switzerland}

\author{Arne Ludwig}
\affiliation{Lehrstuhl f\"{u}r Angewandte Festk\"{o}rperphysik, Ruhr-Universit\"{a}t Bochum, Universit\"{a}tsstr. 150 , D-44780 Bochum, Germany }
\author{Andreas D. Wieck}
\affiliation{Lehrstuhl f\"{u}r Angewandte Festk\"{o}rperphysik, Ruhr-Universit\"{a}t Bochum, Universit\"{a}tsstr. 150 , D-44780 Bochum, Germany }

\author{Daniel Loss}
\affiliation{Department of Physics, University of Basel, Klingelbergstrasse 82, CH-4056 Basel, Switzerland}

\author{Richard J. Warburton}
\affiliation{Department of Physics, University of Basel, Klingelbergstrasse 82, CH-4056 Basel, Switzerland}

\date{\today}

\begin{abstract}
We report a large g-factor tunability of a single hole spin in an InGaAs quantum dot via an electric field. The magnetic field lies in the in-plane direction $x$, the direction required for a coherent hole spin. The electrical field lies along the growth direction $z$ and is changed over a large range, 100 kV/cm. Both electron and hole g-factors are determined by high resolution laser spectroscopy with resonance fluorescence detection. This, along with the low electrical-noise environment, gives very high quality experimental results. The hole g-factor $g_{h}^{x}$ depends linearly on the electric field $F_z$, $dg_{h}^{x}/dF_{z}=(8.3 \pm 1.2)\cdot 10^{-4}$ cm/kV, whereas the electron g-factor $g_{e}^{x}$ is independent of electric field, $dg_{e}^{x}/dF_z=(0.1 \pm 0.3)\cdot 10^{-4}$ cm/kV (results averaged over a number of quantum dots). The dependence of $g_{h}^{x}$ on $F_z$ is well reproduced by a $4 \times 4$ k$\cdot$p model demonstrating that the electric field sensitivity arises from a combination of soft hole confining potential, an In concentration gradient and a strong dependence of material parameters on In concentration. The electric field sensitivity of the hole spin can be exploited for electrically-driven hole spin rotations via the g-tensor modulation technique and based on these results, a hole spin  coupling as large as $\sim 1$ GHz is expected to be envisaged. 
\end{abstract}


\maketitle 

\section{Introduction}
\label{sec:Introduction}
A single electron spin in a self-assembled quantum dot (QD) is a promising candidate for a solid-state qubit \cite{Warburton2013}. In particular, the large optical dipole enables the electron spin to be initialized \cite{Atatuere2006,Gerardot2008}, manipulated \cite{Press2008,Press2010} and read-out \cite{Vamivakas2010,Delteil2014} using fast optical techniques. However, the coupling of the electron spin to the nuclear spin bath of the QD via the hyperfine interaction leads to rapid spin dephasing \cite{Merkulov2002,Khaetskii2002,Warburton2013}. A hole spin represents an alternative \cite{Warburton2013,Gerardot2008,Brunner2009,DeGreve2011}. For a heavy-hole spin, the coefficient describing the hyperfine interaction is about one tenth that of the electron spin \cite{Fischer2008,Fallahi2010,Chekhovich2011}, and, owing to the spin $\pm \frac{3}{2}$ Bloch states, highly anisotropic such that dephasing via the nuclear spins can be suppressed with an in-plane magnetic field \cite{Fischer2008,Testelin2009}. Hole spin dephasing times $T_{2}^{*}$ in InGaAs QDs in excess of 100 ns have been measured in small in-plane magnetic fields \cite{Brunner2009,Houel2014} (although they appear to be smaller at high magnetic fields \cite{DeGreve2011,Greilich2011}) and the decoherence time $T_{2}$ is in the microsecond regime \cite{Brunner2009,DeGreve2011}. The possibilities for manipulating the hole spin in a self-assembled quantum dot electrically have been explored theoretically \cite{Bulaev2007} but not yet experimentally.

We demonstrate here that the hole g-factor in a quantum dot is very sensitive to an electric field $F$ (along the growth direction, $z$) when the magnetic field $B$ is applied in-plane, the magnetic field direction required to generate a coherent hole spin. On the one hand, the sensitivity to electric field implies that charge noise results in hole spin dephasing \cite{Greilich2011,Houel2014}. However, with quiet electrical devices, for instance the ones used here, this limitation can be overcome. On the other hand, the result opens a powerful way to fast electrical control of the hole spin by the g-tensor modulation technique \cite{Salis2001,Kato2003}: the $x$- and $z$-dependencies are different. The predicted hole spin coupling via ac electric field modulation of the g-tensor with a SiGe quantum dot is $\sim 100$ MHz \cite{Ares2013}. Even larger couplings are predicted based on the results presented here.

Recently there have been theoretical \cite{Pryor2006,Pingenot2008,vanBree2012} as well as experimental \cite{Jovanov2011,Bennett2013} studies concerning the g-factor tensor and its tunability for InGaAs QDs. In this paper we augment the measurement methods for determining the g-factor by using resonant laser spectroscopy with resonance fluorescence (RF) detection \cite{Kuhlmann2013,KuhlmannRSI2013}. The method has higher resolution than the detection of photoluminescence following non-resonant excitation. Furthermore, non-resonant excitation introduces not only electrical noise \cite{Houel2012,Kuhlmann2013} and hole spin dephasing \cite{Houel2014} but also creates space charge which screens the applied electric field. These problems are resolved with purely resonant excitation. We are able to combine our high resolution resonance fluorescence experiment with a k$\cdot$p theory to support our experimental results. The k$\cdot$p analysis demonstrates that the origin of the large dependence of $g_{h}^{x}$ on $F_z$ arises from the soft hole confinement potential (allowing the ``center of gravity" of the hole spin wave function to shift in a vertical electric field), an indium concentration gradient (the effective hole ``composition" depends on electric field), and a strong dependence of the material parameters (notably the Luttinger parameter $\kappa$) on indium concentration. 

This paper is organized as follows: In Sec.~\ref{Experiment} we present the experimental setup, the sample design and the measurement technique. The data analysis and the resulting g-factor tunability are discussed in Sec.~\ref{g factor tunability}. In Sec.~\ref{Theory} the experimental results are compared with the theoretical model. For future spin manipulation experiments the realistic Rabi frequencies are estimated in Sec.~\ref{g-tensor modulation}. Followed by Sec.~\ref{Conclusion} with conclusions and final remarks. The derivation of the theory is described in the appendix.\\

\section{Experiment}
\label{Experiment}
\begin{figure}[t]
\includegraphics[width=\linewidth]{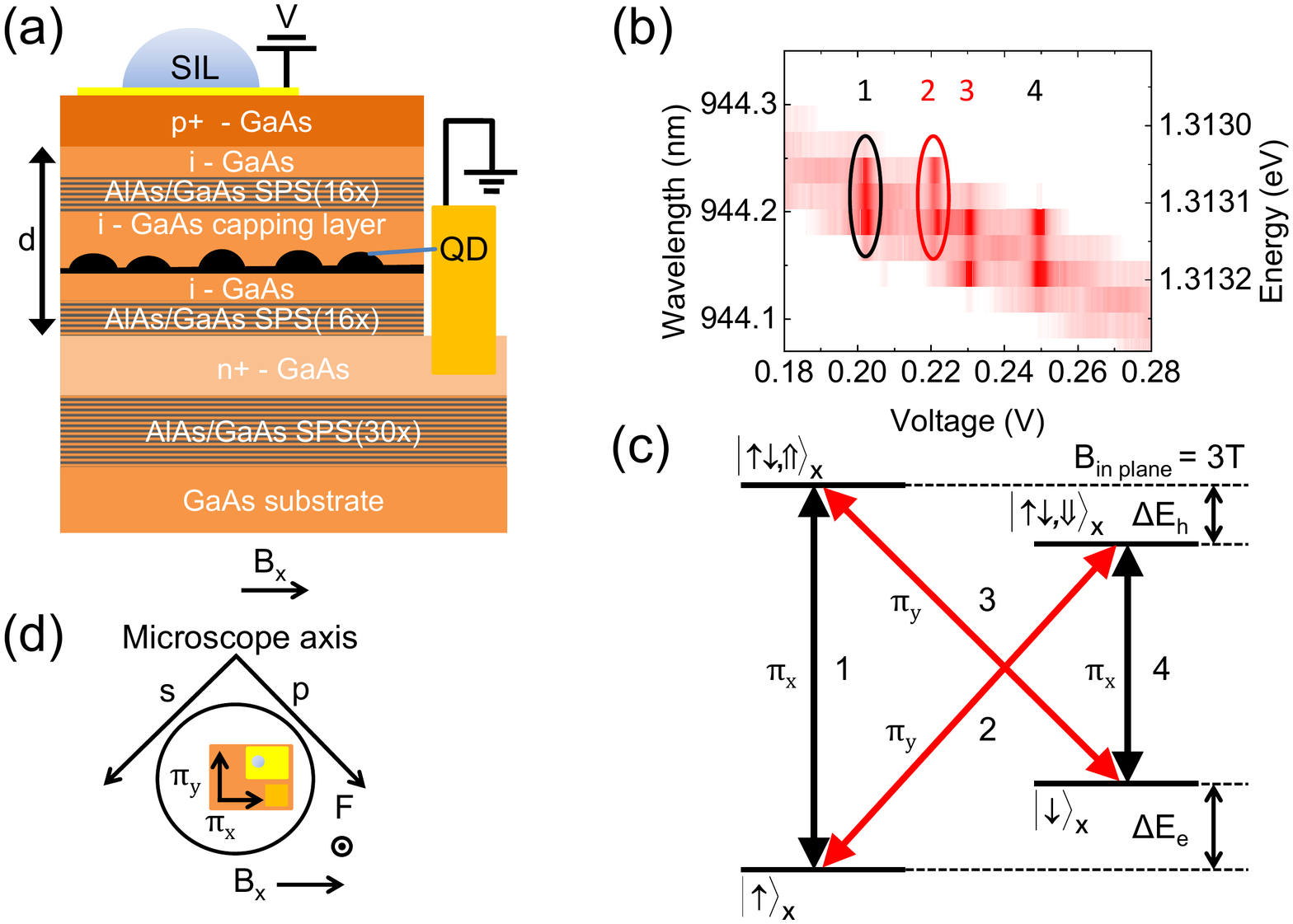}
\vspace*{-0.5cm}
\caption{
(a) Layer sequence of the device. On top of the substrate a short-period superlattice (SPS) of AlAs/GaAs periods is grown, followed by an n$^+$-doped layer, an intrinsic region with a second SPS, the quantum dot layer and a third SPS, completed with a p$^+$ doped layer on top. On the sample surface a semi-transparent electrode is fabricated, with a hemispherical solid-immersion lens (SIL) positioned on top.
(b) Contour plot of the resonance fluorescence (RF) signal as a function of the applied voltage. 1-4 label the four transitions of the charged exciton $X^{1-}$ in an in-plane magnetic field. The transitions are indicated in (c). The color scale is a linear representation of the CCD camera output from background counts (white) to maximum counts 1,200 cts/s (red).
(c) The quantum states of a single electron spin in an in-plane magnetic field. $\uparrow$,$\downarrow$ indicate an electron spin, $\Uparrow$,$\Downarrow$ a hole spin.
(d) Schematic of the sample orientation in the microscope and of the applied fields.
}
\label{fig:1}
\end{figure}

The quantum dots are embedded in the intrinsic region of a p-i-n device. The intrinsic region consists of a layer of self-assembled InGaAs quantum dots located between two highly opaque blocking barriers, in each case an AlAs/GaAs short-period superlattice (16 periods of AlAs/GaAs 3 nm/1 nm).  An electric field $F_z$ of more than 120 kV/cm can be applied to the QDs \citep{Bennett2010}. An etch as deep as the n$^{+}$ layer is followed by annealing Ni/Ge/Au in order to contact the n$^{+}$ GaAs; 60 nm of Au deposited directly onto the surface  without annealing makes a reasonable contact to the p$^{+}$ GaAs. The n$^{+}$ layer is earthed and the electric field is controlled by applying voltage $V$ to the top Au layer, Fig.\ \ref{fig:1}(a). A split-coil magnet inside a He bath cryostat (4.2 K) provides a magnetic field of 3.00 T in the in-plane direction.

\begin{figure*}[t]
\includegraphics[width=\linewidth]{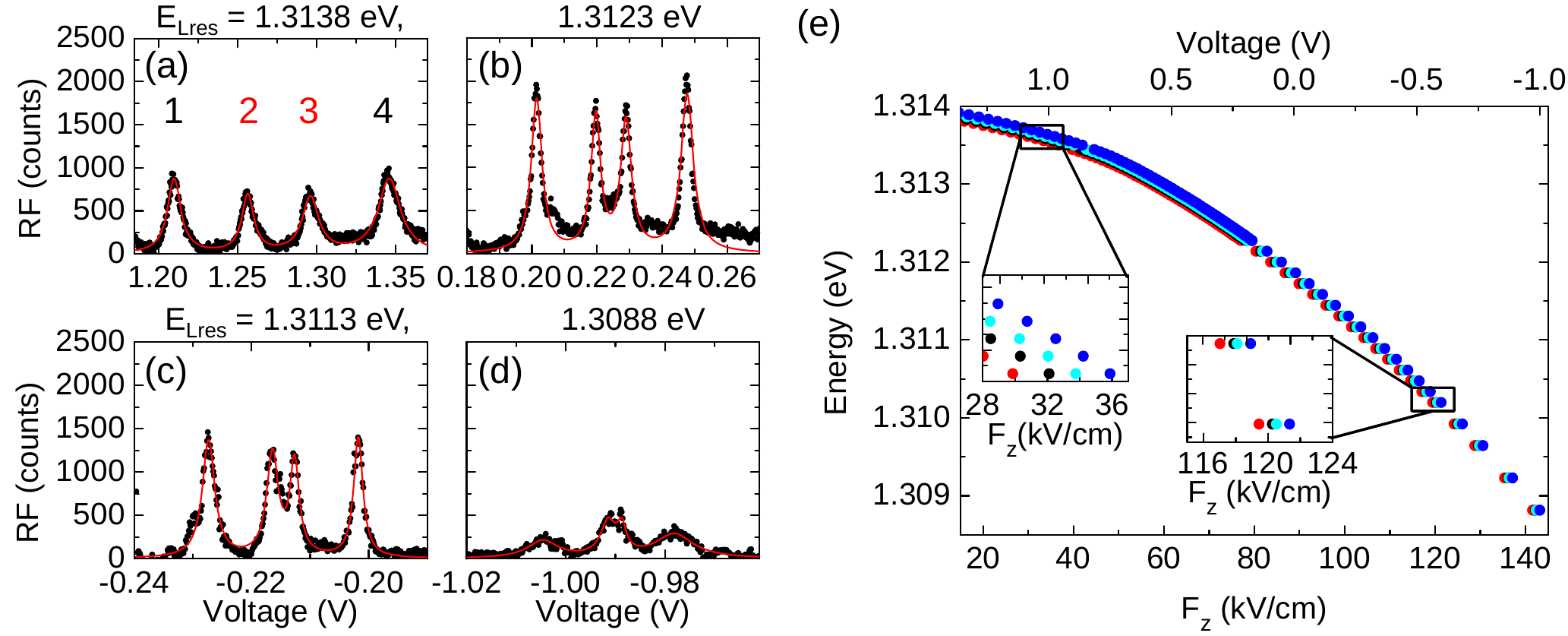}
\vspace*{-0.50cm}
\caption{
(a)-(d) Resonance fluorescence (RF) spectra of the exciton transitions of QD1 at four fixed resonant laser energies. Detuning is achieved by tuning the voltage applied to the device. The 4 spectra are taken at different electric fields spanning the working area of the device and at an in-plane magnetic field of 3.00 T. Each of the four peaks corresponds to one of the four transitions in the quantum system (1-4, Fig.\ \ref{fig:1}(c)). With the help of a quadruple Lorentzian function (red) the peak positions can be determined with a precision of $\pm 0.2$ $\mu$eV. (e) All RF peak positions (resonant excitation energy) versus electric field. The energies can be described in each case by a quadratic function of electric field consistent with the dc-Stark effect. The insets, both with a field extent of 9 kV/cm and an energy extent of 50 $\mu$eV, highlight the electric field dependence of the energy splittings.
}
\label{fig:2}
\end{figure*}

Our experimental scheme involves measuring the frequencies of the optical resonances on single QDs with high resolution laser spectroscopy. We drive the optical transitions with a coherent continuous wave laser (linewidth 1 MHz), collecting the (anti-bunched) resonance fluorescence (RF). The RF is separated from reflected and scattered laser light by a polarization-based dark-field technique \cite{KuhlmannRSI2013}. RF detection is carried out with a charge-coupled device (CCD) at the output of a grating-based spectrometer (resolution $\sim 40$ $\mu$eV). Tuning is carried out by sweeping the transitions through the constant frequency laser, exploiting the dc Stark shift (dependence of QD optical frequency on vertical electric field). The typical linewidths are $\sim 5$ $\mu$eV and in the spectra presented here, we can determine the peak positions with a resolution of $\pm 0.2$ $\mu$eV. We study the negatively charged exciton, the $X^{1-}$. This is advantageous with respect to the neutral exciton, $X^{0}$, in that it exhibits no additional fine-structure splitting due to the electron-hole exchange interaction  \cite{Bayer2002b}: the trion spectrum gives direct access to the electron and hole Zeeman energies, and hence g-factors. The device does not operate in the Coulomb blockade regime where the charge is precisely controlled. However, we find that $X^{1-}$ dominates the optical spectrum in the presence of a small amount of non resonant laser light ($P_{\rm NR}= 0.75$ nW). The in-plane magnetic field along $x$ creates a ``double" $\lambda$-system: the spin-split ground states are both coupled to the spin-split optically-excited states. The ``vertical" transitions in \mbox{Fig.\ \ref{fig:1}(c)} are linearly-polarized along $x$; the ``diagonal" transitions are linearly-polarized along $y$. $x$ corresponds to the [100] crystal direction. The laser is polarized along the ``microscope axis" (s/p) and this corresponds closely to $\pi/4$ with respect to the $x$-axis such that the Rabi couplings of all four transitions are similar. 

Fig.\ \ref{fig:1}(b) shows a contour plot of the RF signal, a plot of RF versus wavelength and $V$. The applied voltage  (electric field) is scanned in 0.2 mV (0.011 kV/cm) steps; the maximum count rate is 1,200 Hz in this case. Depending on the voltage, always two transitions emit together (1,3) and (2,4). This is the experimental signature of the ``double" $\lambda$-system. If for instance the resonant laser drives the ``1"-transition ($\pi_x$-polarized), spontaneous emission takes place via the ``1" recombination channel and also via the ``3" recombination channel ($\pi_y$-polarized). In Fig.\ \ref{fig:1}(b) the RF peaks are assigned to the corresponding energy transitions. Energy separations between peaks 1 and 3, likewise 2 and 4, determine the electron Zeeman energy; energy separations between 1 and 2, likewise 3 and 4, determine the hole Zeeman energy. These energy separations are measured at different electric fields. The applied voltage $V$ is converted into an electric field $F_z$ by calculating the energy band diagram of the entire p-i-n device with a one-dimensional Poisson solver \cite{Snyder3}. To a very good approximation, $F_z=(-V+V_{bi})/d$ where $d$ is the width of the intrinsic region and $V_{bi}$ = 1.52 V is the built in potential. A positive $F_z$ points in the positive $z$ direction.

We note that the experiment does not determine the sign of the g-factors. We make the safe assumption that the electron g-factor is negative: this conjecture has been proven on similar quantum dots emitting at a similar wavelength  \cite{Hoegele2005,Kroner2008c}. With this assumption, we can determine that the hole g-factor has a positive sign. We note also that the hole Zeeman energy is defined as $g_h \mu_B B$ where $\mu_B$ is the Bohr magneton. (This description assigns a pseudo spin of $\pm\frac{1}{2}$ to the hole spin.) 

\section{Electric field dependence of \lowercase{g}-factor}
\label{g factor tunability}
\begin{figure}
\includegraphics[width=\linewidth]{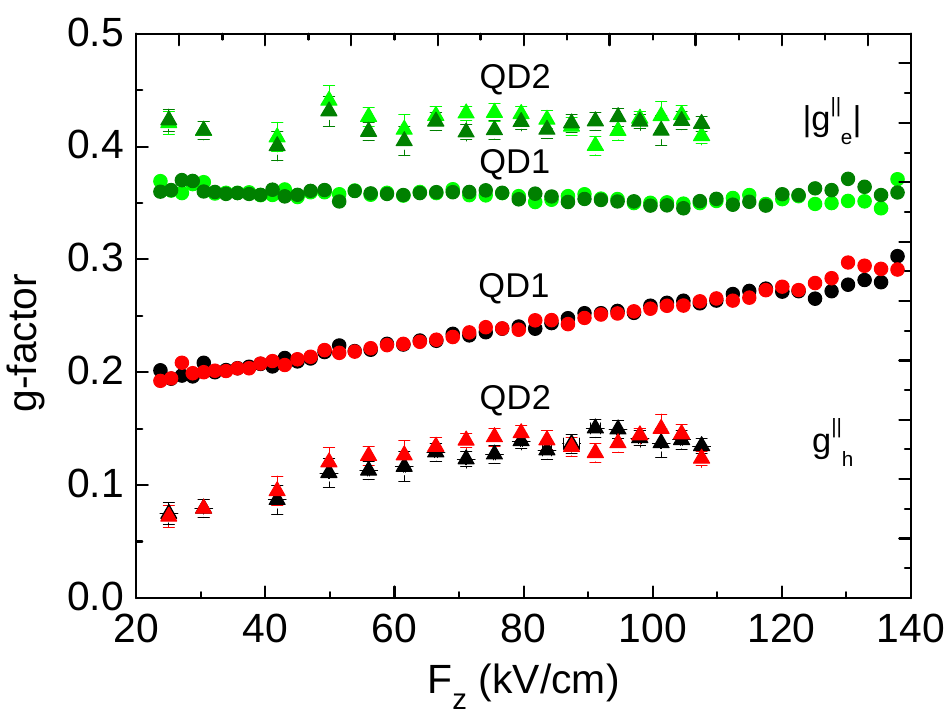}
\vspace*{-0.5cm}
\caption{
In-plane electron $g_{e}^{x}$ and hole $g_{h}^{x}$ g-factors of two QDs as a function of a perpendicular electric field $F_z$. The electron g-factor (green and olive, $E_1 - E_3$ and $E_2 - E_4$), with a mean value of $g_{e}^{x} = -0.39 \pm 0.03$ is not influenced by the external electric field. The in-plane hole g-factor (red and black, $E_1 - E_2$ and $E_3 - E_4$) can be tuned with voltage at a rate of $dg_{h}^{x}/dF_z=(8.3 \pm 0.1) \cdot 10^{-4}$ cm/kV for QD1 (circle) and $dg_{h}^{x}/dF_z=(7.9 \pm 0.9) \cdot 10^{-4}$ cm/kV for QD2 (triangle).
}
\label{fig:3}
\end{figure}

An analysis of the RF recorded at different laser energies (and therefore at different electric fields) reveals voltage, and thus electric field, tuning of the hole g-factor. Four examples are illustrated in Fig.\ \ref{fig:2} (a-d), with a laser energy at the end (a), in the middle (b,c) and at the beginning (d) of the $X^{1-}$ emission energy range. Each RF-spectrum consists of four peaks corresponding to the four transitions of the quantum system (1-4). A quadruple Lorentzian function (red) is used to fit each peak and to determine the peak position. The ``fingerprint" of the spectrum changes from (a) to (d) showing immediately that there is a strong change in the g-factors. To convert the RF spectra versus $V$ into a plot of $g_{h}^{x}$ and $g_{e}^{x}$ versus $V$, we work with Fig.\ \ref{fig:2}(e), a plot of all resonance peak positions, $E_1$, $E_2$, $E_3$ and $E_4$ as a function of $F_z$. Each energy is fitted to a quadratic function of $F_z$, $E = E_0 - p F_z + \beta F_z^2$ \cite{Warburton2002b}. This corresponds to the behavior of an electric dipole with permanent dipole moment $p$ and polarizability $\beta$ in an applied electric field. We find here $p/e=(0.033\pm 0.002)$ nm and $\beta=-(0.234 \pm 0.002)$ $\mu$eV/(kV/cm)$^{2}$ for all four transitions. These values are compatible with the ones reported for similar devices \cite{Bennett2010b}. With this relation it is possible to extract $E_1$, $E_2$, $E_3$ and $E_4$ at all $F_z$. At a particular $F_z$, the energy differences enable us to make a precise measurement of the electron and the hole g-factors. With the relations $g_{h}^{x} \mu_B B = E_1 - E_2 = E_3 - E_4$ and $g_{e}^{x} \mu_B B = E_1 - E_3 = E_2 - E_4$ we determine the in-plane hole and electron g-factors, $g_{h}^{x}$  and $g_{e}^{x}$, for two chosen quantum dots at different in-plane magnetic fields, \mbox{Fig.\ \ref{fig:3}}. 

We find that the electron g-factor $g_{e}^{x}$ does not depend on the applied electric field within the sensitivity of the experiment, $dg_e/dF_z = (0.1\pm 0.3)\cdot 10^{-4}$ cm/kV, and has an mean value of $g_{e}^{x} = -0.39 \pm 0.03$ in our sample. This is true for both quantum dots and magnetic fields up to 3.00 T. The hole g-factor $g_{h}^{x}$ behaves completely differently: there is an approximately linear dependence on $F_z$. The data in Fig.\ \ref{fig:3} show that $g_{h}^{x}$ of QD2 can be tuned by as much as 40\% by changing the applied electric field by 120 kV/cm. A noteworthy fact is that $g_{h}^{x}$ at any one field is highly QD-dependent yet the dependence on field, $dg_{h}^{x}/dF_z$, only changes slightly ($\sim$ 10\%) from QD to QD. We find $dg_{h}^{x}/dF_z=(8.3 \pm 1.2)\cdot 10^{-4}$ cm/kV, a value consistent with previous estimates/measurements \cite{Godden2012b,Bennett2013,Houel2014} where $g_{h}^{x}$ is extracted from the energy splitting of polarization-dependent photoluminescence \cite{Bennett2013,Houel2014} and time-resolved photo-current measurements \cite{Godden2012b}.

\begin{figure}[t]
\includegraphics[width=\linewidth]{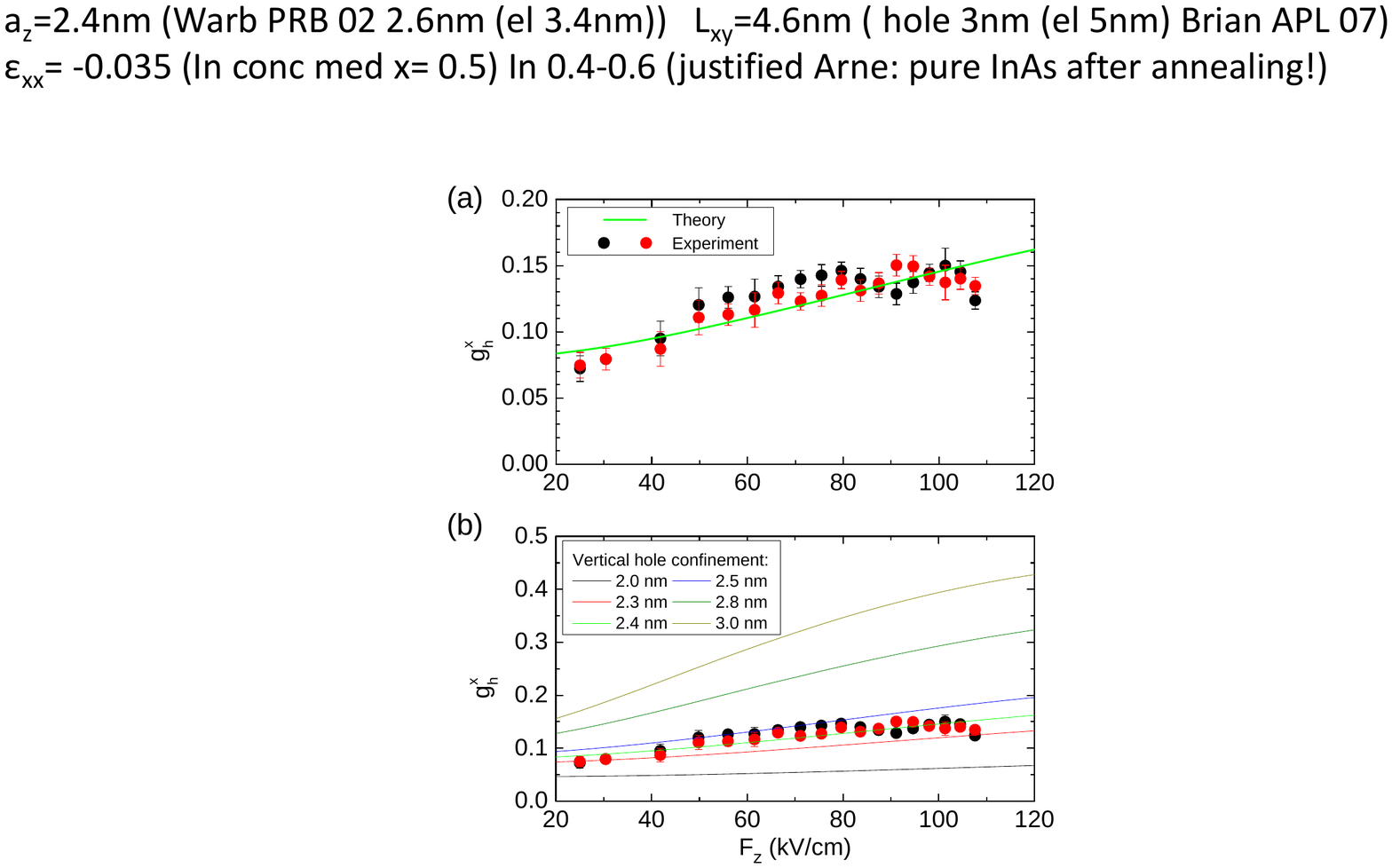}
\vspace*{-0.5cm}
\caption{
(a) The in-plane hole g-factor for QD2 at 3.00 T. For a vertical hole confinement length of 2.4 nm the theoretical description matches the experimental data (solid green line).
(b) Theoretical analysis of the electric field dependence of the in-plane hole g-factor for different values of the vertical hole confinement length.
}
\label{fig:4}
\end{figure}

\section{Theory}
\label{Theory}
The electric field displaces the electron and the hole wave functions inside the QD, by values up to about 0.6 nm in this experiment. The electron wave function is more delocalized than the hole largely on account of its smaller effective mass. The electron averages over a relatively large extent and this averaging does not change significantly on displacing the electron wave function.
This is likely to be the explanation for the small value of $dg_e/dF_z$ observed experimentally  \cite{Pingenot2011}. However, the hole is more strongly localized and even sub-nm displacements have important consequences for $g_{h}^{x}$. An important point is that the QDs have an indium concentration gradient \cite{Mlinar2009}. It is well known that one of the key parameters describing the hole g-factor, the Luttinger $\kappa$-parameter, is highly dependent on indium concentration, changing from for 1.1 GaAs to 7.6 for InAs \cite{Traynor1995}. In the simplest approximation, a pure heavy-hole state has a zero in-plane g-factor: the spin is locked to the angular momentum vector which lies in the $z$ direction by the strong spin-orbit interaction \cite{Martin1990}. However, both the quantum dot confinement potential and the in-plane magnetic field admix the heavy-hole and light-hole states such that there is no simple result for $g_{h}^{x}$.  

Calculations with a four-band k$\cdot$p theory include all the most important sources of heavy-hole, light-hole admixture and provide a quantitative explanation of our experimental results. The confinement induces coupling between the heavy-hole and light-hole states. Strain fields are of considerable strength and are taken into account. The indium content in the quantum dot is assumed to be 40\% at the bottom and 60\% at the top. The external electric and magnetic fields are included perturbatively and the g-factor is derived from the lowest, Zeeman-split hole states. Technical details are described in the Appendix. In Fig.\ \ref{fig:4}(a),(b) the calculated hole g-factor in a magnetic field of 3.00 T is shown as a function of vertical electric field. The results in Fig.\ \ref{fig:4}(b) show that $dg_{h}^{x}/dF_z$ is a strong function of the vertical confinement length of the hole and therefore the height of the quantum dot. 
Independent of material considerations, the lack of useful quantization limits the maximum confinement length, taken as 5 nm here. Significantly, for a realistic quantum dot confinement length of a few nm, the calculations are in good agreement with the experimental results. In fact agreement is excellent for a confinement length of 2.4 nm which is realistic for these QDs (green line), Fig.\ \ref{fig:4}(a). 

\section{\lowercase{g}-tensor modulation}
\label{g-tensor modulation}
Additional measurements in Faraday geometry (magnetic field in growth direction) complete the picture of the hole g-factor tensor. The values we extract are $g_{h}^{z}=1.22\pm 0.02$ at zero electric field for the hole g-factor with an electric field dependence of $dg_{h}^{z}/dF_z=(4.1\pm 1.0)\cdot 10^{-3}$ cm/kV. A similar slope can be found also in previous reports \cite{Jovanov2011}. The dependency of the g-tensor on electric field allows the spin-up and spin-down states to be coupled by applying an ac voltage $V_{\rm ac}$ with a frequency equal to the Larmor frequency. The resulting coupling $f_R$ \cite{Ares2013} is, 
\begin{eqnarray}
f_{R} &=& \frac{\mu_B V_ {\rm ac}}{2 h} \left[\frac{1}{g_{\parallel}} \left(\frac{\partial g_{\parallel}}{\partial V_{g}}\right) 
- \frac{1}{g_{\perp}} \left(\frac{\partial g_{\perp}}{\partial V_{g}}\right)\right] \nonumber\\
&& \times \frac{g_{\parallel}g_{\perp}B_{\parallel}B_{\perp}}{\sqrt{\big(g_{\parallel}B_{\parallel}\big)^2+\big(g_{\perp}B_{\perp}\big)^2}}.
\end{eqnarray}
We estimate a value for the Rabi frequency $f_{R}$ based on our results for the hole g-factor. We consider an oscillating voltage of 1 V (67 kV/cm)
 and a maximum driving frequency of $f_{\rm Larmor}$= 20 GHz. The spin rotation is fastest when the magnetic field is applied at the ``magic" angle, in this case at 20.7$^\circ$ (QD1) and 14.5$^\circ$ (QD2) with respect to the $(x,y)$ plane. The total magnetic field corresponds to 3.2 (4.6) T. These parameters are very reasonable in the sense that the magnetic field is not particularly high and that it lies predominantly in-plane, as required to decouple the heavy-hole spin from the nuclear spins. We obtain resulting Rabi frequencies of 350 MHz (QD1) and 1.1 GHz (QD2), the result depending strongly on the magnitude of the in-plane hole g-factor $g_{h}^{x}$. These very promising values exceed the ones reported for electrons in InSb nanowires \cite{vandenBerg2013} and holes in SiGe QDs \cite{Ares2013}. 

\section{Conclusion and Outlook}
\label{Conclusion}
We have shown the tunability of the hole g-factor of an optically active QD in the key geometry of an in-plane magnetic field (to suppress the coupling of the hole spin to the nuclear spins) and a vertical electric field (experimentally straightforward to apply large fields, here up to 100 kV/cm). We derived the hole and electron g-factors by laser spectroscopy with resonance fluorescence detection. Within the resolution of the experiment, the electron g-factor is independent of the vertical electric field. Conversely, the in-plane hole g-factor is strongly dependent on the vertical electric field with $dg_{h}^{x}/dF_z=(8.3 \pm 1.2)\cdot 10^{-4}$ cm/kV. This result is explained quantitatively with a theoretical model which describes heavy-hole light-hole admixture. The origin of the strong electric field dependence arises from a combination of the softness of hole confining potential, an indium concentration gradient and a strong dependence of material parameters on indium concentration.

A quantum dot hole spin becomes coherent in an in-plane magnetic field. On the one hand, the large $dg_{h}^{x}/dF_z$ implies that charge noise results in hole spin dephasing. This can be minimized of course by working in the clean-material, low-temperature, resonant-excitation limit \cite{Kuhlmann2013}. Another option, as shown by the theoretical calculations, is to work with shallow quantum dots for which $dg_{h}^{x}/dF_z$ is small. On the other hand, the large $dg_{h}^{x}/dF_z$ is useful: g-tensor modulation via an ac electric field can be used to drive spin rotations. With the present quantum dots we predict that the spin can be rotated at frequencies up to $\sim 1$ GHz. An overriding point is that the calculations show the overwhelming influence of the dot height on $dg_{h}^{x}/dF_z$, a powerful route to designing the hole spin properties according to the application.

\begin{acknowledgments}
We thank Christoph Kloeffel for helpful discussions. This work has been supported by the Swiss National Science Foundation (SNF), NCCR QSIT, and IARPA. A.L. and A.D.W. acknowledge gratefully support of Mercur  Pr-2013-0001, BMBF-Q.com-H 16KIS0109, and the DFH/UFA CDFA-05-06.
\end{acknowledgments}

\section*{Appendix: Theory}
\label{Theory app}
We derive the in-plane g-factor of the lowest valence states in a self-assembled InGaAs quantum dot (QD) with an In concentration gradient with applied fields, a vertical electric field and an in-plane magnetic field.  The heavy-hole (HH) and light-hole (LH) states of the bulk material are well described by the $4\times4$ Luttinger Hamiltonian. The strain fields in self-assembled QDs are of considerable strength and affect the band splitting. Strain is therefore incorporated via the Bir-Pikus Hamiltonian. To go from a bulk description to a quantum dot, we add three-dimensional harmonic confinement leading to a change from bands to quantized levels and a mixing of the HH and LH states. We include external out-of-plane electric and in-plane magnetic fields and derive an effective Hamiltonian for the two lowest, Zeeman-split HH states by decoupling them perturbatively from the higher energy states. This effective Hamiltonian is diagonalized exactly allowing the g-factor of this subsystem to be determined. The exact value of $g_{h}^{x}$ depends on the electric field-dependent hole position and the associated local alloy composition within the QD. 

\subsubsection{Hamiltonian}
The Hamiltonians can all be found in Ref.~\cite{Winkler2003}. They are written in terms of the spin-$3/2$ matrices $J_i$, $i=x,y,z$, which are given in a basis of angular momentum eigenstates $\ket{j,m_j}$ with $j=3/2$ and $m_j = \{3/2, 1/2, -1/2, -3/2\}$. Here, the HH band corresponds to $m_j=\pm3/2$ and the LH band to $m_j=\pm1/2$. For our calculations, we locate the origin of the coordinate system at the center of the QD and let the $z$ axis point along the growth direction [001].

The bulk valence band states are described by the Luttinger Hamiltonian
\begin{eqnarray}
H_k&=& -\frac{\hbar^2}{2 m_0}\left[\gamma_1 k^2 -2 \gamma_2 \left[\left(J_x^2-\frac{1}{3}J^2\right)k_x^2+\mathrm{cp}\right]\right]\nonumber\\
&&+\frac{\hbar^2}{2 m_0}4\gamma_3\left[\{J_x, J_y\}\{k_x, k_y\}+\mathrm{cp}\right]\nonumber\\
&& +  \frac{2}{\sqrt{3}}C_k \left[\{J_x, J_y^2-J_z^2\}k_x+\mathrm{cp}\right],
\end{eqnarray}
where  $\{A,B\} = (AB+BA)/2$, cp denotes cyclic permutation, $\hbar k_i = -i \hbar \partial_i$, $i=x,y,z$, is the momentum operator, $k^2 = k_x^2+k_y^2+k_z^2$ and $J^2 = J_x^2+J_y^2+J_z^2$. The $\gamma_l$, $l=1,2,3$, are the Luttinger parameters and the parameter $C_k$ arises as a consequence of the spin-orbit interaction with higher bands. We denote the diagonal part of $H_k$ by $H_{k,0}$. We account for strain by taking into account the Bir-Pikus Hamiltonian
\begin{eqnarray}
H_{{\varepsilon}} &=& D_d \,\mathrm{Tr}\varepsilon+\frac{2}{3}D_u \left[\left(J_x^2-\frac{1}{3}J^2\right)\varepsilon_{xx}+\mathrm{cp}\right]\nonumber\\
&& + \left[C_4 (\varepsilon_{yy}-\varepsilon_{zz})J_x k_x +\mathrm{cp}\right],
\end{eqnarray}
where we consider only diagonal elements $\varepsilon_{ii}$, $i=x,y,z$, of the strain tensor $\varepsilon$ since the off-diagonal shear strain components are negligible everywhere except at the dot interfaces \cite{Tadic2002}. $D_d$ and $D_u$ denote vector potentials and the constant $C_4$ is defined in Ref.~[\onlinecite{Trebin1979}]. In the following, we refer to the diagonal, $\bm{k}$-independent part of $H_{{\varepsilon}}$ as $H_{{\varepsilon},0}$.

We model a flat, cylindrical QD by choosing a harmonic confinement potential, 
\begin{equation}
V_{c} = \left(\begin{array}{cccc}
  V_{c,\text{HH}} &0&0&0\\
  0& V_{c,\text{LH}} &0&0\\
  0&0& V_{c,\text{LH}}&0\\
  0&0&0& V_{c,\text{HH}}
          \end{array}
\right), 
\end{equation}
where
\begin{equation}
V_{c,j}(\mathbf{r}) = -\frac{1}{2}m_{j,\perp} \omega_{j,\perp}^2 z^2 - \frac{1}{2} m_{j,\|} \omega_{j,\|}^2(x^2+y^2),
\end{equation}
with band index $j=\text{HH}, \text{LH}$. The in-plane and out-of-plane confinement energies $\omega_{j, \|}=\hbar/(m_{j, \|}L^2)$ and $\omega_{j, \perp}=\hbar/(m_{j, \perp}a^2)$ are defined by the confinement lengths $L$ and $a$. The corresponding effective masses in the single bands are given by $m_{\text{HH/LH}, \perp}= m_0/(\gamma_1 \mp 2 \gamma_2)$ and $m_{\text{HH/LH}, \|}= m_0/(\gamma_1 \pm \gamma_2)$. We include an external electric field in $z$ direction, $\bm{F} = (0,0,F_z)$, by adding the electric potential
\begin{equation}
 V_{el}(z) = e F_z z. 
\end{equation}
The in-plane magnetic field, $\bm{B} = \nabla \times \bm{A} = (B_x, 0, 0)$, is included by adding two more terms to the Hamiltonian \cite{Ivchenko1998,Kiselev1998}. The first term is found by replacing $\bm{k}\rightarrow\bm{k}+e\bm{A}$ in $H_k+H_{\varepsilon}$ in a semi-classical manner. This yields the implicit magnetic field dependence given by the vector potential $\bm{A}$. We keep only terms linear in $\bm{A}$ and define
\begin{equation}
H_{mc} = e \bm{A} \cdot \bm{v},
\end{equation}
where $\bm{v} = \partial (H_k+H_{\varepsilon})/\partial\bm{k}$ is the velocity operator. We note that proper operator ordering is still enforced. The second term is the magnetic interaction term
\begin{eqnarray}
H_{B}&=& -2 \mu_B [\kappa \bm{J}\cdot\bm{B}+q \bm{\mathcal{J}}\cdot\bm{B}],
\end{eqnarray}
where $\kappa$ is the isotropic and $q$ the anisotropic part of the hole g-factor, $\bm{J} = (J_x, J_y, J_z)$ and $\bm{\mathcal{J}} = (J_x^3, J_y^3, J_z^3)$.

The QD states are then described by
\begin{equation}
H_{qd} = H_k+H_{\varepsilon}+ V_{c} +V_{el}+H_{mc}+H_B. 
\end{equation}
We subdivide $H_{qd}$ into a leading order term 
\begin{equation}
H_{qd,0} = H_{k,0}+H_{{\varepsilon},0}+ V_{c} +V_{el} 
\end{equation}
and a perturbation $H_{qd,1}$. The Hamiltonian $H_{k,0}+ V_{c} +V_{el} $ can be directly mapped onto a three-dimensional, anisotropic harmonic oscillator with an energy shift and a coordinate shift along $z$, both introduced by $V_{el}$. The eigenenergies $E_{j}$ in band $j$ are given by
\begin{eqnarray}
E_{j} &=& \frac{1}{2}\frac{(F_z e)^2}{m_{j,\perp}\omega_{j,\perp}^2}-\hbar \omega_{j,\perp} (n_z+\frac{1}{2})\nonumber\\
&&-\hbar \omega_{j,\|} (n_x+n_y+1).
\end{eqnarray}
The associated eigenfunctions are the usual three dimensional harmonic oscillator eigenfunctions \cite{[{For a textbook example see e.g. }]Sakurai1994} $\phi_{j,\bm{n}}(x,y,z_j)$, where $\bm{n} = (n_x,n_y,n_z)$ is a vector of the associated quantum numbers and $z_j = z-{F_z e}/({m_{j,\perp} \omega_{j,\perp}^2})$. We choose the basis states of $H_{qd,0}$ to be products of type $\phi_{j,\bm{n}}(x,y,z_j)\ket{j,m_j}$. We rewrite $H_{qd}$ in terms of these new basis states and obtain $H_{qd,\text{ext}}$. 

\subsubsection{g-factor}
We are interested in the Zeeman splitting of the two lowest HH states, $\phi_{\text{HH},\bm{0}}\ket{3/2, 3/2}$ and $\phi_{\text{HH},\bm{0}}\ket{3/2, -3/2}$. These states are decoupled from the higher energy states in $H_{qd,\text{ext}}$ by a Schrieffer-Wolff transformation (SWT) of the form $\tilde{H}_{qd,\text{ext}} = e^{-S}H_{qd,\text{ext}}e^{S}$, where $S = -S^{\dagger}$ is an anti-Hermitian operator. The exact procedure is described in detail e.g.\ in Ref.~\onlinecite{Winkler2003}. We perform the SWT up to second order and, by projecting on $\{\phi_{\text{HH},\bm{0}}\ket{3/2, 3/2},\phi_{\text{HH},\bm{0}}\ket{3/2, -3/2}\}$, we obtain an effective, $2\times2$ Hamiltonian $H_{\text{eff}}$. The single elements of $H_{\text{eff}}$ turn out to be too lengthy to be written down here explicitly. Exact diagonalization of $H_{\text{eff}}$ gives two eigenenergies, $E_{\Uparrow}$ and $E_{\Downarrow}$, from which we calculate $g$ according to
\begin{equation}
g_{h}^{x} = \frac{E_{\Uparrow}-E_{\Downarrow}}{\mu_B |\bm{B}|}. \label{eq:gfactor}
\end{equation}

\subsubsection{Hole Position and Material Parameters}
The applied electric field $\bm{F}$ shifts the hole position within the QD along $z$.  Since the QD has an In concentration gradient in the growth direction, the hole experiences an electric-field dependent local material composition. A linear interpolation of the InAs and GaAs material parameters is insufficient to describe ternary alloys. Instead, the gap energy and other band parameters such as the HH mass along [001] and $\kappa$ are given by a quadratic form \cite{Traynor1995,Vurgaftman2001}, where a bowing parameter is introduced to represent the deviation from a linear dependence on composition. We take into account the bowing of the HH mass along $z$ and calculate the hole position within the QD as a function of the applied electric field, $z_{\text{HH}}(F_z)$. This is carried out by minimizing the parabolic part of the HH confinement potential in $z$ direction, 
\begin{equation}
\frac{1}{2} m_{\text{HH},\perp}\omega_{\text{HH},\perp}^2\left[z-\frac{e F_z}{m_{\text{HH},\perp}\omega_{\text{HH},\perp}^2}\right]^2.
\end{equation}
We express the local material composition in terms of the hole position and model the material parameters as functions of $z_{\text{HH}}(F_z)$. Inserting these material parameters in Eq.(\ref{eq:gfactor}) results in $g=g(F_z)$. We observe that the slope of $g(F_z)$ depends strongly on the confinement length $a$, a smaller $a$ corresponding to a flatter QD and less admixture of the LH states to the effective HH states. This effect can be exploited to tailor the observed electric field dependence of $g$ by choosing an appropriate QD height. The confinement lengths are taken according to the values measured on very similar quantum dots, $a = 2.4\mbox{ {nm}}$ \cite{Warburton2002b} and $L = 4.6\mbox{ {nm}}$ \cite{Gerardot2007}. We assumed for In-flushed QDs the In content to be $\sim$ 40\% at the bottom and $\sim$ 60\% at the top of the QD. The average In concentration in combination with the strain parameters for quantum wells ($\varepsilon_{xx}=a_0(\rm{GaAs})/a_0(\rm{In}_{0.5}\rm{Ga}_{0.5}\rm {As})-1$) lead to an estimated strain $\varepsilon_{xx}=\varepsilon_{yy}=-\varepsilon_{zz}=-0.035$ of the system. The material parameters (see Table \ref{tab:parameters}) were modified by the corresponding bowing parameters \cite{Traynor1995,Vurgaftman2001} where available. Note that the values of $q$ reported in the literature \cite{Winkler2003,Lawaetz1971,Glasberg1999,*Toft2007} vary e.g.\ for GaAs between $q_{\text{GaAs}}=0.01-0.04$, meaning that, dependent on $q$, different choices of strain distribution, QD geometry and In profile may produce the same curve. 
\begin{table}[h]
\begin{tabular}{lcllclldd}
\cline{1-4}\cline{6-9}\\[-0.41cm]\cline{1-4}\cline{6-9}
		&& \hspace{-0.15cm}\text{GaAs}	& \text{InAs}	&\mbox{\hspace{0.1cm}}& 	&& \text{GaAs\hspace{-0.3cm}} & \text{InAs\hspace{-0.3cm}}\\	\cline{1-4}\cline{6-9}\\[-0.43cm]
$\kappa$	&& 1.1\text{\cite{Traynor1995}}	& \hspace{0.16cm}7.68\text{\cite{Traynor1995}}	&& $C_k$ &[eV\AA]	& -0.0034  & -0.0112 \\
$q$		&& 0.01\text{\cite{Mayer1991}}	& \hspace{0.16cm}0.04\text{\cite{Lawaetz1971}}	&& $D_d$ &[eV]	& -1.16\text{\cite{Vurgaftman2001}}  & -1.0\text{\cite{Vurgaftman2001}} \\
$\gamma_1$	&& 6.85	& 20.40	&& $D_u$ &[eV]	& 3.0 & 2.7 \\
$\gamma_2$	&& 2.10	& \hspace{0.16cm}8.30	&& $C_4$ &[eV\AA]	& 6.8 \text{\cite{[{We use the values obtained via the LCAO method in }]  Silver1992}} & 7.0 \text{\cite{Silver1992}}\\
$\gamma_3$	&& 2.90	& \hspace{0.16cm}9.10&&&&&\\\cline{1-4}\cline{6-9}\\[-0.41cm]\cline{1-4}\cline{6-9}
\end{tabular}
\caption{Material parameters used in this work. If not stated otherwise, the parameters were taken from Ref.~\cite{Winkler2003}. \label{tab:parameters}}
\end{table}

\end{document}